\documentclass[aps,preprint,noshowpacs,superscriptaddress,groupedaddress]{revtex4}  
\usepackage{graphicx}       
\usepackage{dcolumn}     
\usepackage{bm}            
\usepackage{amssymb}            
\usepackage{color}
\usepackage{ulem} 
\usepackage{graphicx}       
\usepackage{dcolumn}     
\usepackage{bm}            
\usepackage{amssymb}            
\usepackage{color}
\usepackage{amsmath}
\usepackage{ulem} 
\usepackage [english]{babel}
\usepackage [autostyle, english = american]{csquotes}

\MakeOuterQuote{"}

\begin{document}
\setcounter{tocdepth}{3}
\title{Nanotube mechanical resonators with\\ quality factors of up to 5 million}

\author{J. Moser}
\affiliation{ICFO--Institut de Ciencies Fotoniques, Mediterranean
Technology Park, 08860 Castelldefels (Barcelona), Spain}

\author{A. Eichler}
\thanks{Present address: Department of Physics, ETH Zurich, Schafmattstrasse 16, 8093 Zurich, Switzerland}
\affiliation{ICFO--Institut de Ciencies Fotoniques, Mediterranean
Technology Park, 08860 Castelldefels (Barcelona), Spain}

\author{J. G\"uttinger}
\affiliation{ICFO--Institut de Ciencies Fotoniques, Mediterranean
Technology Park, 08860 Castelldefels (Barcelona), Spain}

\author{M. I. Dykman}
\affiliation{Department of Physics and Astronomy, Michigan State
University, East Lansing, Michigan 48824, USA}

\author{A. Bachtold}
\affiliation{ICFO--Institut de Ciencies Fotoniques, Mediterranean
Technology Park, 08860 Castelldefels (Barcelona), Spain}

\begin{abstract}
\textbf{Carbon nanotube mechanical resonators have attracted
considerable interest because of their small mass, the high
quality of their surface, and the pristine electronic states they
host \cite{mass_sensing,Steele2009,Lassagne2009,Ilani}. However,
their small dimensions result in fragile vibrational states that
are difficult to measure. Here we observe quality factors $Q$ as
high as $5\times10^6$ in ultra-clean nanotube resonators at a
cryostat temperature of 30~mK, where we define $Q$ as the ratio of
the resonant frequency over the linewidth. Measuring such high
quality factors requires both employing an ultra-low noise method
to detect minuscule vibrations rapidly, and carefully reducing the
noise of the electrostatic environment. We observe that the
measured quality factors fluctuate because of fluctuations of the
resonant frequency. The quality factors we measure are record
high; they are comparable to the highest $Q$ reported in
mechanical resonators of much larger size \cite{Adiga,Poot}, a
remarkable result considering that reducing the size of resonators
is usually concomitant with decreasing quality factors. The
combination of ultra-low size and very large $Q$ offers new
opportunities for ultra-sensitive detection schemes and quantum
optomechanical experiments.}
\end{abstract}

\maketitle

In recent years, endeavours to boost quality factors in nano and
micromechanical resonators have been stimulated by the need to
develop innovative approaches to sensing \cite{Roukes_sensing},
signal processing \cite{Yamaguchi} and quantum physics
\cite{Aspelmeyer-Meystre-Schwab}. Strategies to enhance quality
factors have proceeded along three main routes. Firstly, the
quality of the host material has been improved. To this end, new
materials have been employed, such as high tensile stress silicon
nitride membranes \cite{Adiga,Weig} and single crystal diamond
films \cite{Degen}. In addition, surface friction has been lowered
by optimizing fabrication processes and reducing contamination
\cite{Painter_APL}. Secondly, schemes to isolate the resonator
from its surrounding environment have been developed, based on new
resonator layouts \cite{Anetsberger,Chan}, and on optical trapping
of thin membranes and levitated particles \cite{Ni,Gieseler}.
Thirdly, and most straightforwardly, $Q$-factors have been
improved by operating resonators at cryogenic temperatures
\cite{Groblacher}.

Schemes to enhance $Q$-factors in nanotube resonators have focused
on reducing contamination by growing ultra-clean nanotubes, and
cooling resonators down to millikelvin temperatures
\cite{Steele2009}. Even though $Q$-factors, measured from the
linewidth of driven resonances, have been improved up to
$\sim1.5\times10^{5}$, they are still much lower than values
routinely obtained with larger resonators fabricated from bulk
materials using top-down techniques \cite{Poot}. This result is
somewhat disappointing, since the high crystallinity of nanotubes
and their lack of dangling bonds at the surface are expected to
minimize surface friction that limits the $Q$-factor in some
nanomechanical systems \cite{Villanueva_Schmid}.

Here we find that the actual values of the $Q$-factors can be
significantly higher than hitherto appreciated, but that revealing
these values requires to perfect the measurement technique.
Namely, the dynamics of the nanotube has to be captured in a
regime of vanishingly small displacement in order to minimize
nonlinear effects. In addition, noise from the electrostatic
environment has to be reduced as much as possible. Indeed, owing
to the ultra-small mass of nanotubes, this electrostatic noise
affects the frequency of the nanotube resonator enormously and
broadens the mechanical linewidth. All these experimental
requirements make it challenging to unmask the instrinsic
$Q$-factor. Progress in measuring ring-down with nanotubes has
been made \cite{Steele_ringdown}, but it has not revealed higher
$Q$-factor values; one reason might be that the large
displacements in those measurements lead to sizeable
nonlinearities.

The geometry of our device, along with its characterization, are
shown in Fig.~1. The nanotube is contacted by source (S) and drain
(D) electrodes, and is suspended over a trench; at the bottom of
the trench is a local gate electrode (Fig.~1a). In this scheme,
the ultra-clean nanotube is grown by chemical vapour deposition in
the last step of the fabrication process of the resonator, making
it free of fabrication residues. The device is cooled to 30~mK, at
which temperature all the data presented here are taken. We
identify the lowest lying flexural mode of the resonator using the
frequency-modulated (FM) mixing technique
\cite{Gouttenoire,Alex2011}, and study its dependence on gate
voltage $|V_{g}^{DC}|$ (Fig.~1b). Figure~1c shows resonance
lineshapes in response to oscillating electrostatic forces,
measured with the FM technique; they yield $Q=1700$. The voltages
used to produce these driving forces are kept low so that the
resonance lineshapes are just above the noise floor.

In the following, we employ an ultra-sensitive detection method
described in Ref.~\cite{force_sensing} to capture the tiny
amplitude of the thermal vibrations at 30~mK. In contrast to the
FM technique, the resonator is not driven by an oscillating force.
Displacement fluctuations are transduced into current
fluctuations, of which power spectra $S_{I}$ are measured (see
Methods). We operate the resonator within the electron Fabry-Perot
regime (Fig.~1d), where the effect of electron transport on
mechanical vibrations is less pronounced than in the Coulomb
blockade regime (Fig.~1e)
\cite{Steele2009,Lassagne2009,Ilani,Wernsdorfer2012}. A typical
resonance lineshape, obtained by averaging power spectra for
$\sim512$~s, is shown in the upper panel of Fig.~2a. The
corresponding $Q$-factor is $Q\sim4\times10^{5}$ for a resonant
frequency $f_{0}=55.6$~MHz. Remarkably, such a $Q$-factor is 200
times higher than that measured with the FM mixing technique
(Fig.~1c).

We find that both the $Q$-factor and $f_{0}$ fluctuate in time. To
show this, spectra obtained over a measurement time $\tau=3.2$~s
are acquired successively, using either a commercial source by
Keithley or a simple lead battery to supply $V_{g}^{DC}$ (see
colour plots of $S_{I}$ as a function of frequency and time in
Figs.~2a,b). Frequency fluctuations with the Keithley source are
larger than with the lead battery. The $Q$-factor is fluctuating
as well; its distribution is broad and asymmetric (Figs.~2c,d). On
average, $Q$-factors are higher with the lead battery than with
the Keithley source. While fluctuations in $f_{0}$ have been
observed in high-$Q$ mechanical resonators
\cite{Kippenberg_NatureComm,Tang_PRB,Roukes_PRL2013}, fluctuations
in the measured $Q$-factor have not been discussed thus far. We
show below that both fluctuations are closely related.

The averaged $Q$-factor decreases with the measurement time
(Fig.~3a). Once the measurement time is set, the $Q$-factor does
not reveal any marked dependence on the amplitude $V_{sd}^{AC}$ of
the oscillating source-drain bias used to read out current
fluctuations (Fig.~3b). This shows that $V_{sd}^{AC}$ does not
affect the measured motion of the resonator. For example, if the
resonator were to heat up due to Joule heating, the $Q$-factor
would decrease as $V_{sd}^{AC}$ increases; if the resonator were
to self-oscillate, the $Q$-factor would vary as $V_{sd}^{AC}$
increases. Neither the $Q$-factor nor $f_{0}$ changes as
$V_{g}^{DC}$ is swept through a conductance oscillation (see
Supplementary Section VI). This confirms that the effect of
electron transport on the vibrations is weak. This is crucial to
our experiment, since Coulomb blockade would reduce the $Q$-factor
\cite{Steele2009,Lassagne2009,Ilani,Wernsdorfer2012}.

The observed frequency fluctuations are associated with the
electrostatic noise of the environment. In the case of the
Keithley source, frequency fluctuations are traced back to the
voltage fluctuations of the source itself as a result of the
strong gate voltage dependence of $f_{0}$. To show this, we
transform voltage fluctuations into Allan deviation $\sigma_{A}$
using the $f_{0}$-to-$V_{g}^{DC}$ conversion factor measured in
Fig.~1b, $\sim1.4\times10^7$~Hz/V (see Methods). In Fig.~3c, we
find that the measured Allan deviation of the resonant frequency
of the nanotube (green hollow circles) coincides with the Allan
deviation expected from the gate voltage fluctuations generated by
the source (blue trace). The frequency noise is of $1/f$-type
(pink noise), since $\sigma_{A}$ is essentially constant as a
function of measurement time $\tau$. By contrast, the lead battery
is stable enough that it does not significantly affect $f_{0}$.
The Allan deviation expected from the voltage fluctuations of the
lead battery is smaller than $\sigma_{A}(\tau)$ measured with the
nanotube resonator (Fig.~3d). The origin of the observed frequency
fluctuations with the lead battery is not clear. Since
$\sigma_{A}$ remains constant as a function of $\tau$ ($1/f$
noise), $f_{0}$ fluctuations might be related to two-level systems
such as charge fluctuators in the gate dielectric.

The $Q$-factor is affected by the electrostatic noise of the
environment as well. This can be inferred from a comparison
between Fig.~2a and Fig.~2b. Larger fluctuations of the resonant
frequency (lower panels) result in a larger resonance linewidth
(upper panels), hence in a lower $Q$-factor.

The fluctuations of the quality factor, the asymmetry of its
distribution, and its dependence on measurement time are all
attributed to the fluctuations of the resonant frequency. To show
this, we separate this frequency noise into a slow part and a fast
part on the scale of the ring-down time (see Methods and
Supplementary Section X). For a measurement time $\tau$ short
compared to the characteristic time scale of the slow frequency
noise $\tau_{slow}^{c}$, resonances have a Lorentzian lineshape
whose width $2\tilde{\Gamma}$ is constant while its resonant
frequency fluctuates from one measurement to the next. For
$\tau\gg\tau_{slow}^{c}$, resonances are broadened by the time
average of many such Lorentzians, resulting in non-Lorentzian
lineshapes (Fig.~3e). For $\tau\sim\tau_{slow}^{c}$, linewidths
are larger than $2\tilde{\Gamma}$, and fluctuate from one
measurement to the next. Upon increasing $\tau$, linewidths tend
to increase because the resonator has time to explore a larger
frequency range. The distribution of linewidths is asymmetric
since linewidths have a fixed lower bound given by
$2\tilde{\Gamma}$ (see Supplementary Section X).

Figures~2e,f display two high $Q$ resonance lineshapes, obtained
with both gate voltage sources, and demonstrating
$Q=3.5\times10^{6}$ (Keithley source) and $Q=4.8\times10^{6}$
(lead battery). We verify that the areas of the resonances in
$S_{I}$ spectra in Figs.~2e,f are equal (within 10\%) to the areas
of the averaged resonances in Figs.~2a,b. This confirms that these
sharp resonances do capture the dynamics of the nanotube, in spite
of the relatively low number of measurement points dictated by the
short measurement time. A slightly lower quality factor is
obtained in an additional device, also operated in the Fabry-Perot
regime (see Supplementary Section IX), confirming the robustness
of high $Q$ resonances. These $Q$-factors are comparable to the
highest ones measured in large micromechanical resonators
\cite{Adiga}. Yet, they may still be limited by the measurement,
and may not have reached the intrinsic $Q$-factor of nanotube
resonators, defined as $f_{0}$ multiplied by the small-amplitude
ring-down time (the ring-down time is not known, but it is smaller
than $\tau$; see Supplementary Section X).

The large quality factors observed here are associated with the
small amplitude of the nanotube vibrations. Using the
equipartition theorem, we obtain that the mode temperature is
$T=44\pm10$~mK, which corresponds to a phonon population
$n=k_{B}T/hf_{0}=16\pm4$ with $k_{B}$ the Boltzmann constant and
$h$ the Planck constant (see Supplementary Section V). In this
estimate, we use the areas of the resonances in Fig.~2, which are
all equal within 10\%. From the temperature, we calculate that the
variance of displacement is $\sim(35\textrm{ pm})^{2}$.

Larger displacements translate into lower $Q$-factors. This is
illustrated in Fig.~3f where a white voltage noise is applied to
the gate electrode in order to enhance the displacement. Possible
origins of this behaviour could be associated to nonlinear damping
forces, which result in a mechanical linewidth that depends on the
amplitude of motion, and to the spectral broadening of the
resonance, which is induced by the combination of nonlinear
conservative forces and displacement fluctuations
\cite{Alex2011,DKreview84,Alex_Nature_Comm,McEuen_PNAS}. Note that
the applied voltage noise also induces fluctuations of $f_{0}$,
but their contribution to the measured resonance linewidth is
negligible (see Supplementary Section VII). The temperature
dependence of the $Q$-factor could not be measured in our current
measurement setup.

The giant quality factors and the associated weak fluctuations of
$f_{0}$ hold promise for diverse sensing experiments. The limit to
force sensing is ultimately set by the force noise $S_{f}=8\pi
Mk_{B}Tf_{0}/Q$ \cite{force_sensing}. Using
$M=4.4\times10^{-21}$~kg, $Q=4.8\times10^{6}$, and $T\sim44$~mK,
we obtain $\sqrt{S_{f}}\sim10^{-21}$~N$/\sqrt{\textrm{Hz}}$, which
is lower than what has been achieved with mechanical resonators
thus far \cite{force_sensing}. The sensitivity of mass sensing and
force gradient detection is given by the Allan deviation of
frequency fluctuations ($\delta f_{0}\simeq8$~Hz for
$\tau\simeq6$~s) in Fig.~3d obtained with the lead battery. The
latter translates into a mass resolution $\delta M=2M\times\delta
f_{0}/f_{0}\simeq10^{-27}$~kg, which is as good as the best
estimates on record \cite{mass_sensing}. We extract a force
gradient resolution of $8\pi^{2}Mf_{0}\delta
f_{0}\simeq10^{-10}$~N/m; this compares favorably with the best
values reported thus far \cite{Marohn}.

Nanotube resonators are relevant candidates for exploring the
quantum regime of nanomechanics. Their low mass leads to large
zero-point motion, $\sqrt{\hbar/(4\pi f_{0}M)}\simeq6$~pm, and
allows to enhance the coupling to other quantum systems. High
$Q$-factors are necessary for the quantum manipulation of
mechanical states. In an experiment whereby the resonator is
cooled to the ground state with passive feedback, the lifetime of
the ground state is $\tau_{ph}=\hbar Q/k_{B}T\simeq10^{-3}$~s (for
$k_{B}T>hf_{0}$) using $T\sim44$~mK and assuming that
$Q=4.5\times10^{6}$ is related to the ring-down time. Our nanotube
resonator demonstrates $Qf_{0}=3\times10^{14}$~Hz, which compares
well with the highest values measured thus far in nano and
micromechanical resonators \cite{Chan,Laird}.

Overall, our work shows that the $Q$-factor of ultra-clean
nanotube resonators can reach very high values, provided that the
resonators are cooled to low enough temperature for nonlinear
effects to be negligible, that slow frequency noise is reduced,
that measurement times are short, and that the effect of the
coupling between vibrations and Coulomb blockade is suppressed.
The unique combination of ultra-low mass and giant quality factors
offers new opportunities for ultra-sensitive detection schemes and
optomechanical experiments in the quantum regime.
\vspace{20pt}\\
\textbf{Methods}

\textbf{Detecting thermal vibrations.} We employ an
ultra-sensitive detection method that allows us to capture the
tiny amplitude of the thermal vibrations in a dilution
refrigerator cooled to 30~mK. Displacement fluctuations $\delta z$
in the direction $\hat{z}$ normal to the gate are transduced into
conductance fluctuations $\delta
G=\frac{dG}{dV_{g}}V_{g}^{DC}\frac{C_{g}^{\prime}}{C_{g}}\delta
z$, where $C_g$ is the capacitance between the nanotube and the
gate, and $C_{g}^{\prime}=dC_{g}/dz$ (for the estimation of
$C_{g}^{\prime}$, see Supplementary Section II). Large transduced
signals require a large transconductance $dG/dV_{g}$. The largest
transconductance is obtained in the Coulomb blockade regime, for
$V_{g}^{DC}>0$ (Fig.~1e), as is usually observed in ultra-clean
suspended nanotubes \cite{Steele_scientific_report}. However,
Coulomb blockade reduces $Q$-factors, since it amplifies the
coupling between electron transport and mechanical vibrations
\cite{Steele2009,Lassagne2009,Ilani,Wernsdorfer2012}. For this
reason, we operate the resonator within the Fabry-Perot regime,
realized for $V_{g}^{DC}<0$, even though $dG/dV_{g}$ is lower
(Fig.~1d). In this regime, oscillations in conductance originate
from quantum interferences of electronic waves, and the effect of
the coupling between vibrations and Coulomb blockade is weaker.
The tiny conductance fluctuations $\delta G$ are parametrically
down-converted into low frequency current fluctuations $\delta I$
by applying a small oscillating voltage of amplitude $V_{sd}^{AC}$
across source and drain, at a frequency $f_{sd}$ shifted by
$\sim10$~kHz from the mechanical resonant frequency $f_{0}$. Power
spectra $S_{I}$ of fluctuations $\delta I$ are measured at
frequencies $|f_{sd}-f_{0}|\sim10$~kHz with a cross-correlation
technique using a vector signal analyzer \cite{force_sensing}.

\textbf{Allan deviation.} We calculate the Allan deviation
$\sigma_{A}$ as
\begin{equation}
\sigma_{A}^{2}(\tau)=\frac{1}{2(N-1)}\sum_{i=1}^{N-1}\left(\frac{\langle
f_{i+1}\rangle_{\tau}-\langle f_{i}\rangle_{\tau}}{\langle
f_{0}\rangle}\right)^{2}\,,
\end{equation}
where $\langle f_{i+1}\rangle_{\tau}$ and $\langle
f_{i}\rangle_{\tau}$ are two subsequent measurements of $f_{0}$
averaged over the integration time $\tau$, $N$ is the number of
averaged frequency measurements, and $\langle f_{0}\rangle$ is the
average of $f_{0}$ over the whole measurement \cite{Allan}. We
define $f_{0}$ as the frequency for which $S_{I}$ is largest.

\textbf{Power spectrum of displacement and frequency noise.} We
separate the frequency noise into a slow part and a fast part on
the scale of the ring-down time. The power spectrum of
displacement $S_{q}$ (which is proportional to $S_{I}$) as a
function of angular frequency $\omega=2\pi f$ reads:
\begin{equation}
S_{q}(\omega)=\frac{k_{B}T}{M\omega_{0}^{2}}\frac{1}{\tau}\int_{0}^{\tau}dt\frac{\tilde{\Gamma}}{{\tilde{\Gamma}}^{2}+[\omega-\tilde{\omega}_{0}-\xi_{slow}(t)]^{2}}\,,
\label{Sq}
\end{equation}
where $k_{B}$ is the Boltzmann constant, $T$ is the mode
temperature, $t$ is the time, and $M$ is the modal mass.
$\tilde{\Gamma}$ is the sum of the reciprocal ring-down time and
of the broadening due to fast frequency noise,
$\tilde{\omega}_{0}$ is the resonant angular frequency
renormalized by fast frequency noise, and $\xi_{slow}$ is the slow
part of frequency noise (see Supplementary Section X). The area
$\int_{0}^{\infty} S_{q}(\omega)d\omega=\pi
k_{B}T/M\omega_{0}^{2}$ is given by the equipartition theorem,
hence it is independent of frequency noise.

\begingroup

\renewcommand{\addcontentsline}[3]{}
\renewcommand{\section}[2]{}

\endgroup

\vspace{20pt} \textbf{Acknowlegements}\\ We thank H. Flyvbjerg and
S. N{\o}rrelykke for discussions. We acknowledge support from the
European Union through the ERC-carbonNEMS project (279278), a
Marie Curie grant (271938), and the Graphene Flagship, MINECO and
FEDER (MAT2012-31338), the Catalan government (AGAUR, SGR), and
the US Army Research Office.
\vspace{20pt}\\
\textbf{Author contributions}\\ J.M. developed the experimental
setup, carried out the measurements, and analyzed the data. A.E.
fabricated the devices. J.G. provided support with the
experimental setup. M.I.D. and A.B. provided support with the
analysis. M.I.D. wrote Supplementary Section X. J.M., M.I.D. and
A.B. wrote the manuscript with critical comments from all authors.
A.B. and J.M. conceived the experiment. A.B. supervised the work.
\vspace{20pt}\\
\textbf{Additional information}\\
The authors declare no competing financial interests.
Supplementary information accompanies this paper at
www.nature.com/naturenanotechnology. Reprints and permission
information is available online at
http://npg.nature.com/reprintsandpermissions/. Correspondence and
requests for materials should be addressed to AB.
\vspace{20pt}\\

\begin{figure}[!ht]
\begin{center}
\includegraphics{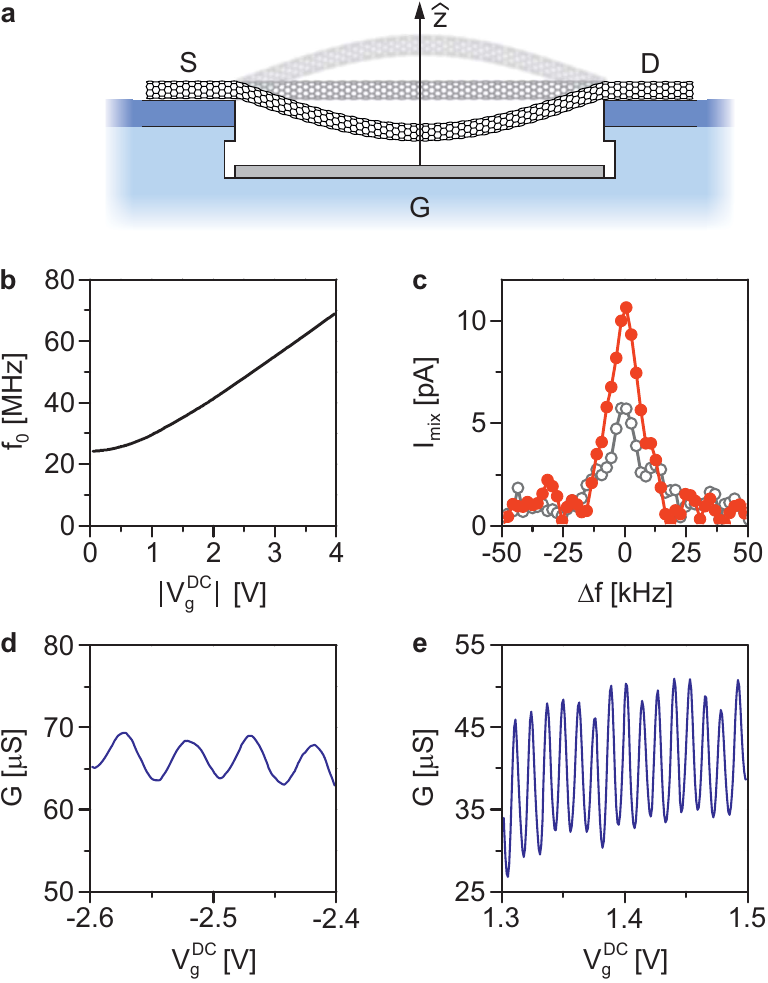}
\caption{\textbf{Carbon nanotube mechanical resonator.} (a)
Schematic of the device. The nanotube is contacted by source (S)
and drain (D) electrode, and is suspended over a gate electrode
(G). The trench has a width of $1.8$~$\mu$m and a depth of
$\sim350$~nm. (b) Resonant frequency $f_{0}$ as a function of gate
voltage $|V_{g}^{DC}|$ for the lowest lying mechanical mode. (c)
Mixing current $I_{mix}$ as a function of drive frequency $\Delta
f$ (measured from $f_{0}=44.1$~MHz), using the FM technique.
$V_{g}^{DC}=-2.2$~V. Two different driving voltage amplitudes are
used: 13~$\mu$V (filled circles) and 8~$\mu$V (hollow circles). We
use a lock-in amplifier with a time constant of 300~ms. In the FM
technique, the $Q$-factor is extracted from the width of the foot
of the resonance \cite{Gouttenoire,Alex2011}, yielding
$Q\simeq1700$. (d-e) Conductance $G$ as a function of
$V_{g}^{DC}$, in the Fabry-Perot regime (d), and in the Coulomb
blockade regime (e). The full $G(V_{g}^{DC})$ trace is shown in
Supplementary Section I.}
\end{center}
\end{figure}

\begin{figure}[!ht]
\includegraphics{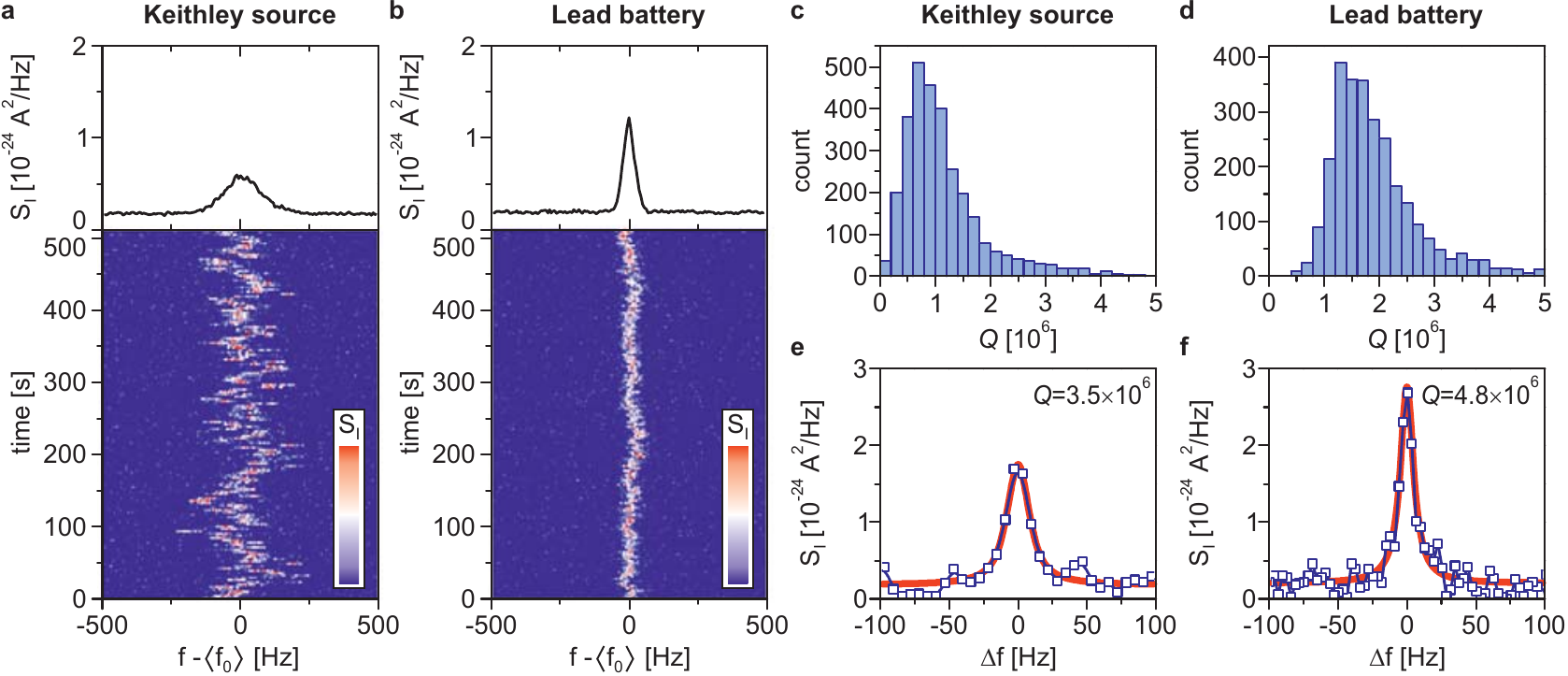}
\caption{\textbf{Fluctuations of the $Q$-factor and of the
resonant frequency.} (a, bottom) Power spectra of current
fluctuations $S_{I}(f-\langle f_{0}\rangle)$ acquired successively
in time, using a commercial DC voltage source (Keithley) to supply
the gate voltage ($V_{g}^{DC}=-3.037$~V). For each of these
spectra, the measurement time is 3.2~s. (a, top) Resonance
obtained by averaging all the spectra. (b) Same experiment as in
(a), this time using a lead battery to supply the gate voltage.
Scale bar: blue, power spectral density
$S_{I}=0.15\times10^{-24}$~A$^{2}$/Hz; red,
$S_{I}=2.50\times10^{-24}$~A$^{2}$/Hz. (c, d) Histograms of
$Q$-factor using the Keithley source (c) and the lead battery (d),
constructed from 3000 power spectra acquired with the same
settings as in (a, b). (e, f) Examples of high $Q$ resonances
obtained with the Keithley source (e) and the lead battery (f).
Red curves are fit to Lorentzian functions. For all panels,
$\langle f_{0}\rangle=55.6$~MHz is the time-averaged resonant
frequency.}
\end{figure}

\begin{figure}[!ht]
\begin{center}
\includegraphics{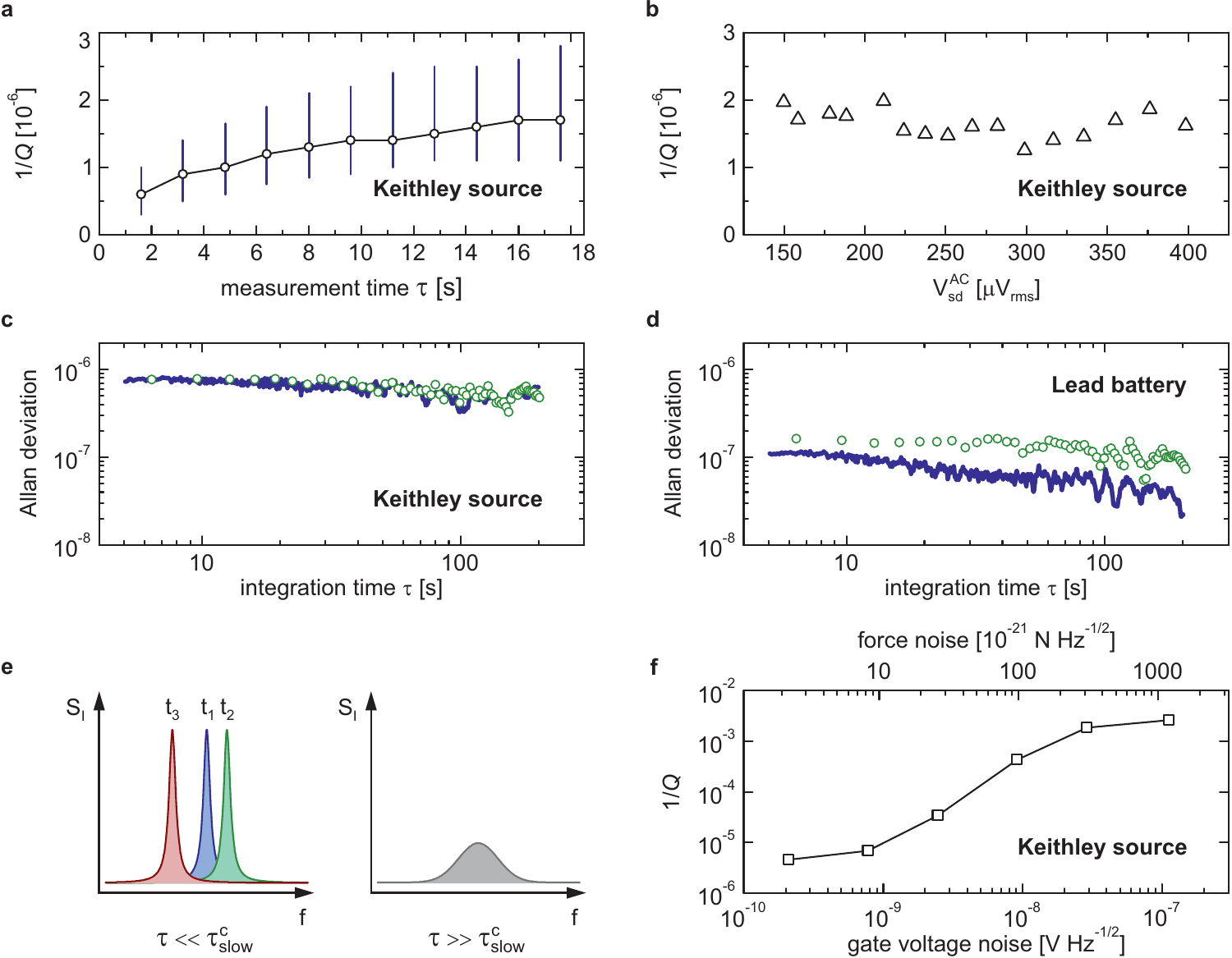}
\caption{\textbf{Characterization of the $Q$-factor and of the
fluctuations in $f_{0}$.} (a) $1/Q$ as a function of measurement
time $\tau$, using the Keithley source. For each measurement time,
300 power spectra are acquired and a histogram of $1/Q$ is built.
Circles denote $1/Q$ at the maximum of the histograms, and bars
represent the full width at half maximum of the histograms. The
resonant frequency is $f_{0}=55.6$~MHz for panels (a-d). (b) $1/Q$
as a function of $V_{sd}^{AC}$ for $\tau=16$~s. (c, d) Allan
deviation of $f_{0}$ as a function of $\tau$ with the Keithley
source (c) and with the lead battery (d). The green circles are
for the measured resonant frequencies, and the blue traces are for
the resonant frequencies estimated from gate voltage fluctuations.
(e) Schematic of resonance lineshapes for short (left) and long
(right) measurement times $\tau$ compared to the characteristic
time of the slow frequency noise $\tau_{slow}^{c}$. $t_{1,2,3}$
are three successive instants. (f) $1/Q$ as a function of added
gate voltage noise at $V_{g}^{DC}=-2.2$~V ($f_{0}=44.1$~MHz) for
$\tau=16$~s. The gate voltage noise is applied using the
Johnson-Nyquist noise of a 50~Ohm resistor at room temperature
amplified by different gains. It creates a white electrostatic
force noise between the nanotube and the gate (top axis).}
\end{center}
\end{figure}

\setcounter{figure}{0}
\setcounter{equation}{0}
\renewcommand{\thefigure}{S\arabic{figure}}
\renewcommand{\theequation}{S\arabic{equation}}
{\center{\Large{\textsc{Supplementary Information}}}}

\tableofcontents

\newpage

\section{Conductance as a function of gate voltage}
\begin{figure}[b]
\includegraphics{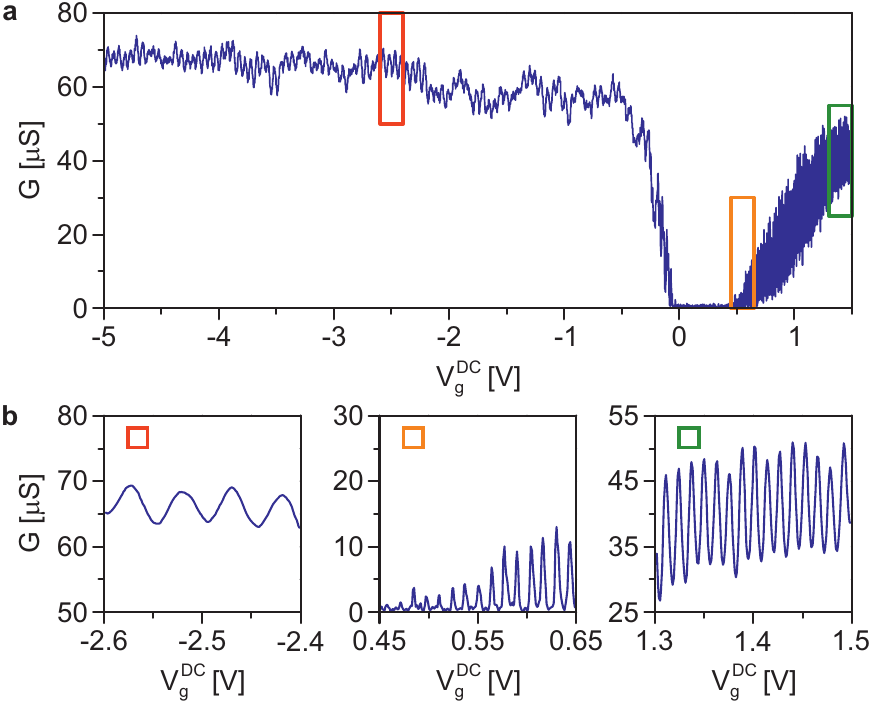}
\caption{(a) Conductance $G$ as a function of gate voltage
$V_{g}^{DC}$. Regions within colored rectangles are blown up in
(b).} \label{conductance}
\end{figure}
The conductance $G$ as a function of gate voltage $V_{g}^{DC}$ at
30~mK is shown in Supplementary Fig.~\ref{conductance}. The regime
for positive $V_{g}^{DC}$ corresponds to Coulomb blockade
(Supplementary Fig.~\ref{conductance}b, middle and right-hand-side
panels). Indeed, the length of the blockaded region, estimated to
be $\sim1.5$~$\mu$m from the separation between conductance
oscillations, is close to the width of the trench (see next
section). For negative $V_{g}^{DC}$, charge transport is in the
so-called Fabry-Perot regime (Supplementary
Fig.~\ref{conductance}b, left-hand-side panel). The period in
$V_{g}^{DC}$ of the oscillations of $G$ is a few times larger than
the period of oscillations for positive $V_{g}^{DC}$, as usually
observed for suspended nanotubes. In addition, the conductance is
close to $2e^{2}/h$, indicating that the electronic transmission
of each contact is high.

The measurement shown in Supplementary Fig.~\ref{conductance} is
typical of ultra-clean nanotubes \cite{Benyamini}. The regions of
the nanotube near the metal electrodes remain $p$-doped, due to
the work function of the electrodes, whereas the suspended part of
the nanotube can be doped with electrons or holes using
$V_{g}^{DC}$. For positive $V_{g}^{DC}$, $p-n$ junctions are
formed near the metal electrodes, resulting in a Coulomb blockaded
region along the suspended nanotube. For negative $V_{g}^{DC}$,
the nanotube is $p$-doped along the whole tube and no tunnel
barriers are formed. Near $V_{g}^{DC}=0$, the conductance is zero
because the Fermi energy lies inside the energy band gap along the
entire suspended part of the nanotube.

\section{Estimating $C_{g}^{\prime}$}
We estimate the capacitance $C_{g}$ between the nanotube and the
gate from the separation $\Delta V_{g}^{DC}=13\pm1$~mV between two
conductance peaks in the Coulomb blockade regime (Supplementary
Fig.~\ref{conductance}b, central and right-hand-side panels):
$C_{g}=e/\Delta V_{g}^{DC}=1.2\pm0.1\times10^{-17}$~F. This value
is close to the model capacitance between a cylinder of length $L$
and radius $r$, and a plane a distance $h$ away:
\[
C_{g}=\frac{2\pi\epsilon_{0}L}{\ln\left(\frac{2(h-z)}{r}\right)}\,,
\]
where $\epsilon_{0}$ is the vacuum permittivity, $h$ is the
separation between the nanotube and the gate, and $z\ll h$ is a
small displacement of the cylinder in the direction normal to the
gate. The derivative $dC_{g}/dz$ reads

\[
C_{g}^{\prime}=\left(\frac{dC_{g}}{dz}\right)_{z=0}=\frac{C_{g}}{h\ln(2h/r)}=\left(5.2\pm1.9\right)\times10^{-12}\textrm{
F/m}\,,
\]
using $h=350\pm50$~nm and $r=1\pm0.5$~nm (the typical radius of
nanotubes obtained with our chemical vapour deposition technique).

\section{Estimating the effective mass}
We study the fundamental flexural mode of the nanotube resonator
(see Fig.~1b of the main text). We use a simple model whereby the
nanotube is straight and is perpendicular to the trench. The
effective mass $M$ of the mode is related to the mass of the
nanotube $M_{NT}$ as
\[
M=M_{NT}\frac{1}{L}\int_{0}^{L}[\phi(x)]^{2}dx\,,
\]
where $L$ is the length of the nanotube and $\phi(x)$ is the shape
of the mode, which is normalized so that $\max[\phi(x)]=1$. Given
that mechanical tension is induced in the nanotube by the contacts
and by the gate voltage, we assume that the modal shape is
$\phi(x)=\sin(\pi x/L)$. The latter is the simplest approximation
for the shape of a beam under tension. We emphasize that the
expression for $M$ takes into account the shape of the vibrational
mode; all other quantities are measured with respect to the
amplitude of this mode. To estimate $M_{NT}$, we also assume that
the length of the nanotube is equal to the trench width
($L=1.8\pm0.2$~$\mu$m) and that $r=1$~nm as in Section II.
Therefore
\[
M=\frac{1}{2}\left(2M_{C}\times\frac{2\pi r\times
L}{A}\right)=\left(4.4\pm2.4\right)\times10^{-21}\textrm{ kg}\,,
\]
where $M_{C}=2\times10^{-26}$~kg is the mass of a carbon atom and
$A=5.2\times10^{-20}$~m$^{2}$ is the surface area of a hexagon in
the honeycomb lattice of graphene.

\section{Orientation of the mode with respect to the gate
electrode}

\begin{figure}[t]
\includegraphics{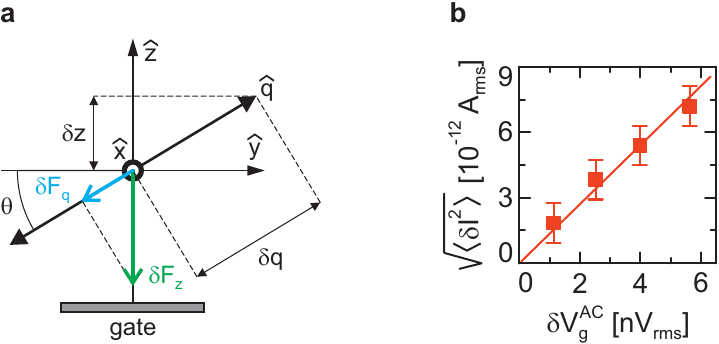}
\caption{(a) Orientation of the displacement for a simple model of
a straight nanotube parallel to the gate electrode. $\hat{x}$ and
$\hat{y}$ are parallel to the gate, and $\hat{z}$ is normal to the
gate. The displacement (that is, the mode polarization) is along
$\hat{q}$. The force $\delta F_{z}$ along $\hat{z}$ acts on the
motion of the nanotube through the force $\delta F_{q}$, its
projection onto $\hat{q}$. The projection of displacement $\delta
q$ onto $\hat{z}$ is $\delta z$. (b) Standard deviation of current
$\sqrt{\langle\delta I^{2}}\rangle$ induced by a small oscillating
gate voltage of amplitude $\delta V_{g}^{AC}$ applied on
resonance. $\delta V_{g}^{AC}$ produces the force $\delta F_{z}$
in (a).} \label{theta}
\end{figure}

The orientation of the vibrations of the eigenmode depends on the
static curvature of the suspended nanotube, which builds in during
the fabrication of the resonator. Since the static curvature
cannot be controlled and cannot be accurately measured, the
eigenmode can vibrate in any direction. For a real nanotube, the
displacement along the nanotube can be quite complicated. However,
we will use a simplified model where at least for the lowest mode
the displacement is in one plane. We will characterize this plane
by the angle $\theta$ it makes with the plane parallel to the
surface of the gate electrode (Supplementary Fig.~\ref{theta}a).

We assume that current fluctuations $\delta I$ at the drain
electrode are proportional to motional fluctuations $\delta z$
along the direction $\hat{z}$ normal to the gate electrode. Then,
the current at the frequency close to the difference between the
mode eigenfrequency and the frequency of the source-drain voltage
is
\begin{equation}
\delta I=\beta\delta
z=\frac{1}{2}\frac{dG}{dV_{g}}V_{g}^{DC}V_{sd}^{AC}\frac{C_{g}^{\prime}}{C_{g}}\delta
z\,, \label{I2z}
\end{equation}
where $dG/dV_{g}$ is the transcondutance, $V_{g}^{DC}$ is the
static gate voltage, and $V_{sd}^{AC}$ is the amplitude of the
oscillating source-drain voltage. We assume that the mode is
polarized along $\hat{q}$, as shown in Supplementary
Fig.~\ref{theta}a; hence, $\delta z$ is the projection along
$\hat{z}$ of the motional fluctuation $\delta q$ along $\hat{q}$:
\begin{equation}
\delta z=\delta q\sin\theta\,, \label{z2q}
\end{equation}
where $\theta$ is the angle between $\hat{y}$ and $\hat{q}$. To
estimate $\theta$, we subject the nanotube to a weak electrostatic
force by applying a small oscillating voltage with amplitude
$\delta V_{g}^{AC}$ to the gate. The component of the force along
$\hat{z}$ is $\delta F_{z}=C_{g}^{\prime}V_{g}^{DC}\delta
V_{g}^{AC}$. The center of mass of the nanotube experiences a
force $\delta F_{q}=\delta F_{z}\sin\theta$, the projection of
$\delta F_{z}$ along $\hat{q}$. On resonance, the displacement
$\delta q$ induced by $\delta F_{q}$ reads
\begin{equation}
\delta q=\frac{Q}{M\omega_{0}^{2}}\delta F_{q}\,,\label{q2Fq}
\end{equation}
where $Q$ is the quality factor of the resonance, $M$ the
effective mass, and $\frac{\omega_{0}}{2\pi}$ the resonant
frequency. In turn, the displacement $\delta q$ induces current
oscillations of variance $\langle\delta I^{2}\rangle$ at the
drain:
\begin{equation}
\sqrt{\langle\delta
I^{2}}\rangle=\beta\sin^{2}\theta\frac{Q}{M\omega_{0}^{2}}C_{g}^{\prime}V_{g}^{DC}\delta
V_{g}^{AC}=p\delta V_{g}^{AC}\,.\label{slope}
\end{equation}
Supplementary Fig.~\ref{theta}b shows $\sqrt{\langle\delta
I^{2}\rangle}$ as a function of $\delta V_{g}^{AC}$, measured with
the technique described in the main text and in
Ref.~\cite{force_sensing}. The dependence of $\sqrt{\langle\delta
I^{2}\rangle}$ on $\delta V_{g}^{AC}$ is linear and extrapolates
to zero in the limit of zero $\delta V_{g}^{AC}$, as expected from
Eq~(\ref{slope}). We obtain $\theta=60^{\circ}\pm25^{\circ}$ from
the slope in Supplementary Fig.~\ref{theta}b, using
$|V_{g}^{DC}|=2.2$~V, $V_{sd}^{AC}=3.17\times10^{-4}$~V,
$C_{g}=(1.2\pm0.1)\times10^{-17}$~F,
$C_{g}^{\prime}=(5.2\pm1.9)\times10^{-12}$~F/m,
$M=(4.4\pm2.4)\times10^{-21}$~kg,
$\omega_{0}/(2\pi)=44.1\times10^{6}$~Hz, $Q=5\times10^{5}$,
$dG/dV_{g}=6.4\times10^{-4}$~S/V, and
$p=(1.3\pm0.2)\times10^{-3}$~A/V.

The estimation of $\theta$ comes with a large uncertainty that
mostly originates from the uncertainties in the estimations of
$C_{g}^{\prime}$ and $M$. These uncertainties, however, have
little impact on the estimation of the modal temperature $T$,
which we present is Section V. Indeed, we show that $T$ is
proportional to $M/(C_{g}^{\prime}\sin\theta)^{2}$, a quantity
that we can estimate rather precisely from the measurement shown
in Supplementary Fig.~2b. For this, we rewrite Eq.~(\ref{slope})
as:
\begin{equation}
\frac{M}{(C_{g}^{\prime}\sin\theta)^{2}}=\frac{1}{p}\left(\frac{1}{2}\frac{dG}{dV_{g}}(V_{g}^{DC})^{2}V_{sd}^{AC}\frac{1}{C_{g}}\frac{Q}{\omega_{0}^{2}}\right)\,.
\label{p}
\end{equation}
Using the values given above, we calculate the right-hand-side of
Eq.~(\ref{p}) and obtain
\[
\frac{M}{(C_{g}^{\prime}\sin\theta)^{2}}=205\pm40\textrm{
kg}\cdot\textrm{m}^{2}/\textrm{F}^{2}\,.
\]
We use this value to estimate $T$ is Section V.

\section{Mode temperature}
Using Eqs.~(\ref{I2z}) and (\ref{z2q}), the equipartition theorem
yields
\begin{equation}
k_{B}T=M\omega_{0}^2\langle\delta
q^{2}\rangle=M\omega_{0}^2\frac{1}{(\beta\sin\theta)^{2}}\frac{1}{2\pi}\int_{-\infty}^{\infty}{S_{I}(\omega)d\omega}\,,
\label{equipartition}
\end{equation}
where $\langle\delta q^{2}\rangle$ is the variance of displacement
along $\hat{q}$, and $S_{I}(\omega)$ is the \textit{two-sided}
power spectral density of current fluctuations $\delta I$. To make
contact with the experiment, we also express $k_{B}T$ in terms of
$S_{I}(2\pi f)$, where $f=\frac{\omega}{2\pi}$ is the natural
frequency:
\begin{equation}
k_{B}T=M\omega_{0}^2\frac{1}{(\beta\sin\theta)^{2}}\int_{0}^{\infty}2S_{I}(2\pi
f)df\,, \label{Tjm}
\end{equation}
where we used
\[
\frac{1}{2\pi}\int_{-\infty}^{\infty}S_{I}(\omega)d\omega=\int_{-\infty}^{\infty}S_{I}(2\pi
f)df=2\int_{0}^{\infty}S_{I}(2\pi f)df\,.
\]
In Eq.~(\ref{Tjm}), $S_{I}(2\pi f)$ is defined as
\begin{equation}
S_{I}(2\pi
f)=\lim_{\tau\rightarrow\infty}\int_{-\tau/2}^{\tau/2}\langle\delta
I(t)\delta I(0)\rangle\exp[-i(2\pi
f+\omega_{sd}-\omega_{0})t]dt\,,
\end{equation}
where $\tau$ is the measurement time, $\omega_{sd}$ is the
off-resonance angular frequency of the source-drain voltage, and
$\omega_{0}$ is the resonant angular frequency. Experimentally, we
measure the \textit{single-sided} power spectral density
$S_{I}^{\textrm{exp}}(2\pi f)=2S_{I}(2\pi f)$.

We estimate the mode temperature $T$ from the power spectra in
Figs.~2e,f of the main text. We obtain $T\simeq44\pm10$~mK, using
the following parameters: $|V_{g}^{DC}|=3.037$~V,
$dG/dV_{g}=8.5\times10^{-4}$~S/V, $V_{sd}^{AC}=4\times10^{-4}$~V,
$C_{g}=(1.2\pm0.1)\times10^{-17}$~F,
$\omega_{0}/(2\pi)=55.6\times10^{6}$~Hz, and
$\int_{0}^{\infty}{S_{I}^{\textrm{exp}}(2\pi
f)df}=4.5\times10^{-23}$~A$^{2}$. We also use
$M/(C_{g}^{\prime}\sin\theta)^{2}=205\pm40\textrm{
kg}\cdot\textrm{m}^{2}/\textrm{F}^{2}$ obtained in Section IV.

We were not able to carry out a temperature dependence of
$\langle\delta z^{2}\rangle$. Connecting thermometry lines to the
cryostat would generate electrical noise in our measurement,
masking the mechanical resonance in the spectra.

\section{Dependences of the resonant frequency and the $Q$-factor on DC gate voltage}

\begin{figure}[t]
\includegraphics{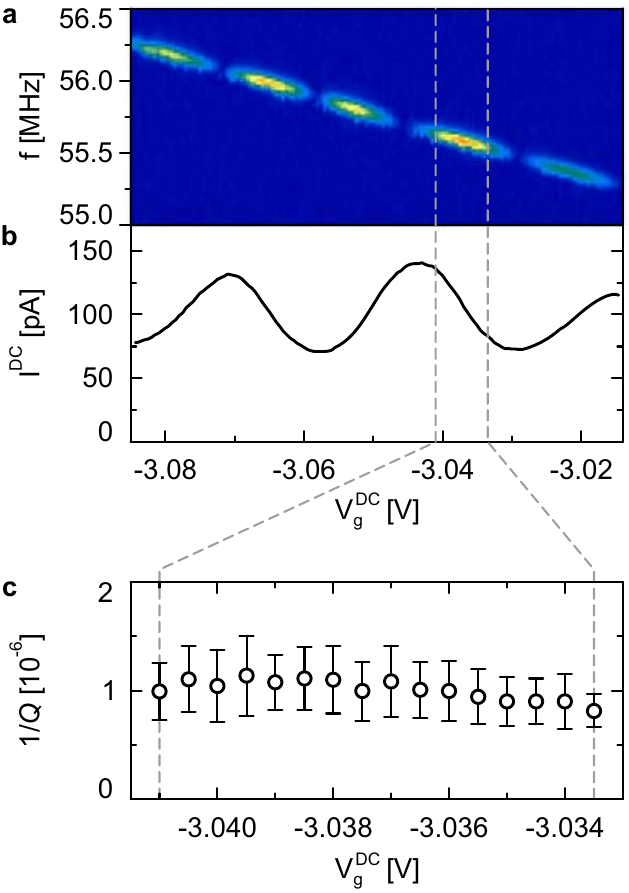}
\caption{(a) Electro-mechanical mixing current measured with the
FM technique, as a function of $V_{g}^{DC}$ and drive frequency
$f$. Scale bar from blue, 1~pA, to orange, 140~pA. (b)
Source-drain DC current $I^{DC}$ as a function of $V_{g}^{DC}$,
showing Fabry-Perrot oscillations. (c) $1/Q$ as a function of time
averaged $V_{g}^{DC}$. Measurement time is 6.3~s. The error bars
correspond to the distribution of $Q$-factors for 20 power
spectra.}\label{Vg}
\end{figure}

Supplementary Fig.~\ref{Vg}a shows the resonant frequency as a
function of $V_{g}^{DC}$ obtained by measuring the mixing current
$I_{mix}$ with the FM technique \cite{Alex2011} as a function of
$V_{g}^{DC}$ and drive frequency $f$. The intensity of $I_{mix}$
varies with $V_{g}^{DC}$ because it is proportional to the
transconductance. The resonant frequency $f_{0}$ is linear in
$V_{g}^{DC}$. Supplementary Fig.~\ref{Vg}c shows the inverse of
the $Q$-factor as $V_{g}^{DC}$ is stepped from the bottom of an
oscillation in $I^{DC}(V_{g}^{DC})$ to the top of this
oscillation. No variation of $1/Q$ is seen.

The dependences of $f_{0}$ and $1/Q$ on $V_{g}^{DC}$ are different
from what is observed in nanotube resonators operated in the
Coulomb blockade regime. There, the resonator experiences a
reduction of $f_{0}$ near the top of an oscillation in $I^{DC}$ as
a function of  $V_{g}^{DC}$. In addition, $1/Q$ is higher near the
top of the oscillation in $I^{DC}$ because electron tunneling
events are accompanied by enhanced dissipation. The absence of
such behaviors demonstrates that the effect of the coupling
between vibrations and Coulomb blockade is weak in our experiment.

\section{Effect of electrostatic white noise on the spectrum of
the resonator.}

Figure~3f of the main text displays $1/Q$ in the presence of gate
voltage noise of various intensities. The gate voltage noise is
applied using the Johnson-Nyquist noise of a 50~Ohm resistor at
room temperature amplified by different gains (by varying the
number of amplifiers). We measure the power spectral density of
the amplified Johnson-Nyquist noise with a signal analyzer in a
separate experiment. We verify that this amplified noise is white
(the power spectral density is constant) between 1~MHz and
200~MHz, a frequency range that encompasses the resonant frequency
of our resonator (40-60~MHz). Noise at frequencies below 1~MHz is
cut off by a high pass filter at the sample stage. The power
spectral densities we measure are consistent with the gain (20~dB)
and the noise figure (1~dB) of each amplifier used in combination
with the total attenuation along our radio frequency lines.

The applied Johnson-Nyquist noise creates a random electrostatic
force between the nanotube and the gate (top axis of Fig.~3f). To
the lowest order in the nanotube displacement, this force is
additive, which means that it is independent of the displacement.
The nanotube resonator responds primarily to the frequency
components of this force, which lie within the band centered at
the resonant frequency, with the typical bandwidth given by the
mechanical linewidth. The power spectrum of the random force is
flat in this region, and therefore the corresponding noise is
white. Its intensity $S_{ef}$ is related to the power spectral
density of the Johnson-Nyquist noise $S_{JN}$ as $S_{ef} =
(C_{g}^{\prime}V_g^{ DC}\sin \theta)^2 S_{JN}$ (see Section IV).
We use $C_{g}^{\prime}= 5.2\times 10^{-12}$~F/m, $V_g^{ DC} =$~2.2
V, and $\theta = 60^\circ$.

The effect of the additive Johnson-Nyquist noise is fully
analogous to the effect of the thermal noise that comes along with
the resonator dissipation. It leads to the increase of the
intensity of the peak in the resonator power spectrum and to
spectral broadening of this peak via nonlinear damping and via the
dependence of the oscillator frequency on the vibration amplitude.

The low-frequency components of the Johnson-Nyquist noise could
directly lead to fluctuations of the resonator frequency, which
would be similar to the low-frequency fluctuations induced by the
voltage noise of the DC voltage source. They would come from the
dependence of the resonator frequency on the gate voltage. Such
noise is multiplicative, since the corresponding force is
proportional to the resonator displacement, and this is why it
acts as a frequency shift. The resonator responds to such noise if
its frequencies are low, within the typical bandwidth given by the
mechanical linewidth. Higher-frequency components are averaged
out. For the Johnson-Nyquist noise that we are using the effect of
this noise is very weak, because the initial noise from the 50~Ohm
resistor is only weakly amplified below 100~kHz, and is further
attenuated by a high-pass filter below 10~kHz. We estimate that
the corresponding gate voltage fluctuations induce a lineshape
broadening of only a fraction of a Hertz, much smaller than the
linewidths we measure in Fig.~3f.

\section{Voltage fluctuations of the DC voltage sources}

\begin{figure}[t]
\includegraphics{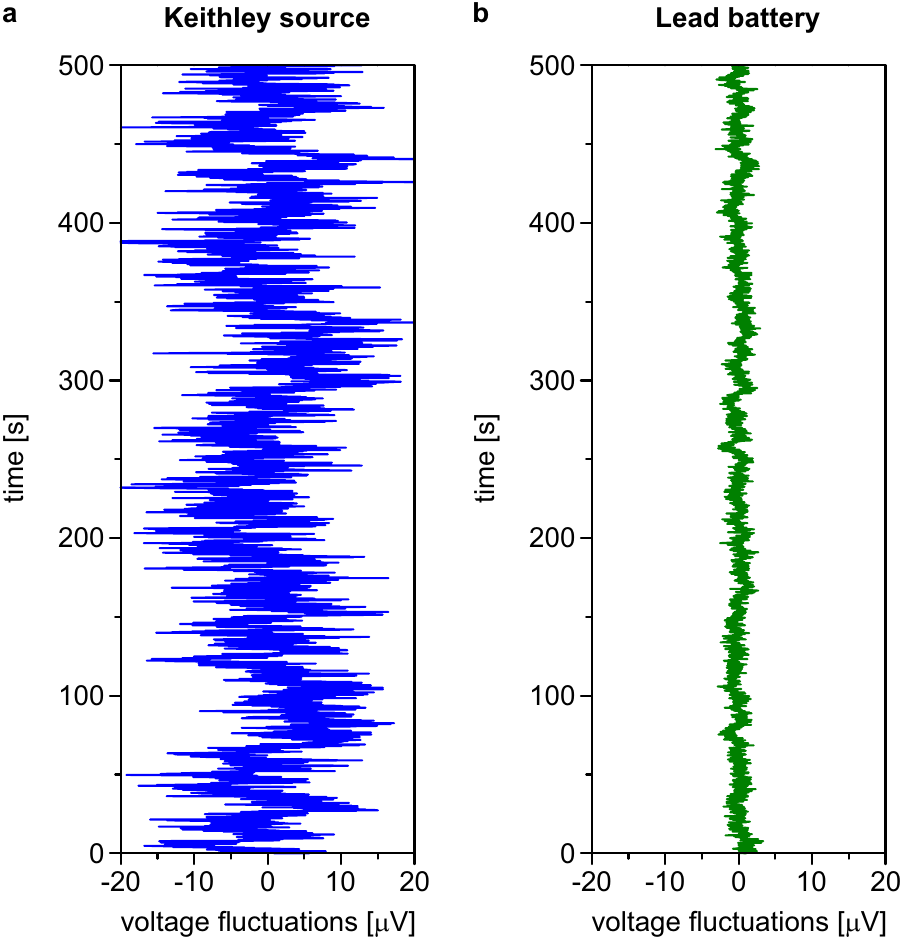}
\caption{Gate voltage fluctuations measured at the sample stage.
(a) Keithley source. (b) Lead battery.} \label{voltage_noise}
\end{figure}

The voltage fluctuations of both the Keithley source and of the
lead battery are measured in a separate experiment. The DC source
is connected to the DC gate voltage input port outside of the
cryostat. A finite DC voltage close to the gate voltages used in
the main text is applied. The voltage fluctuations at 300~K are
measured at the sample stage after opening the cryostat. These
fluctuations are recorded with a voltmeter as a function of time
(Supplementary Fig.~\ref{voltage_noise}). The measured voltage
fluctuations are larger than the intrinsic readout fluctuations of
the voltmeter. These measurements are used to estimate the Allan
deviation in Figs.~3c,d (blue traces) of the main text.

\section{Additional sample}
We investigated a second nanotube resonator device operated in the
Fabry-Perot regime at a refrigerator temperature of 30~mK. An
example of a high $Q$ resonance for this device is shown in
Supplementary Fig.~\ref{secondsample}. The parameters used for
this measurement are: $V_{g}^{DC}=-3.643$~V,
$V_{sd}^{AC}=4\times10^{-4}$~V, and a measurement time of 4.8~s.
The measurements were carried out with the Keithley source.
\begin{figure}[h]
\includegraphics{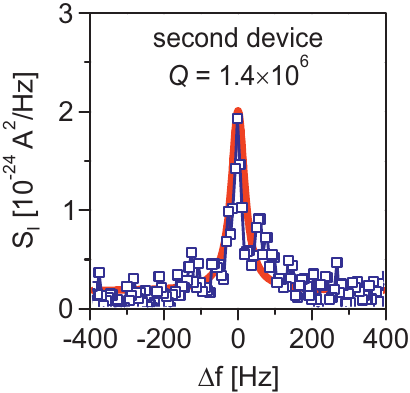}
\caption{High $Q$ resonance obtained with an additional device.
The resonant frequency is $f_{0}\simeq62\times10^{6}$~Hz.}
\label{secondsample}
\end{figure}

\section{Power spectrum for a finite measurement time}
Below, power spectral densities are all \textit{two-sided}. The
power spectrum of displacement fluctuations for a finite
measurement time $\tau$ reads
\begin{equation}
S_{q}(\omega)=\frac{1}{\tau}\int_{0}^{\tau}dt\int_{0}^{\tau}dt_{1}\,q(t)q(t_{1})e^{i\omega(t-t_{1})}=\frac{2}{\tau}\textrm{Re}\int_{0}^{\tau}dt\int_{0}^{t}dt_{1}\,q(t)q(t_{1})e^{i\omega(t-t_{1})}\,.
\label{Sq}
\end{equation}
We make the change from fast oscillating variables $q(t)$,
$\dot{q}(t)$ to slow complex oscillator amplitude $u(t)$
\begin{equation}
\begin{split}
q(t)&=u(t)e^{i\omega_{0}t}+\textrm{c.c.}\,,\\
\dot{q}(t)&=i\omega_{0}\left(ue^{i\omega_{0}t}-u^{\ast}e^{-i\omega_{0}t}\right)\,.
\end{split}
\label{qdq}
\end{equation}
Plugging Eqs.~(\ref{qdq}) into the equation of motion
$\ddot{q}+2\Gamma\dot{q}+[\omega_{0}^{2}+2\omega_{0}\xi(t)]q=f_{T}(t)/M$,
and solving for $u$ in the rotating wave approximation, we obtain
\begin{equation}
u(t)=\int_{-\infty}^{t}dt^{\prime}\,\exp\left[-\Gamma(t-t^{\prime})+i\int_{t^{\prime}}^{t}\xi(t^{\prime\prime})dt^{\prime\prime}\right]f_{u}(t^{\prime})\,.\label{u}
\end{equation}
In the above expression, $\xi$ is the frequency noise,
$f_{u}=1/(2iM\omega_{0})f_{T}(t)\exp(-i\omega_{0}t)$ is the
thermal noise with correlator $\langle
f_{u}(t)f_{{u}^{\ast}}(t^{\prime})\rangle=\frac{\Gamma
k_{B}T}{M\omega_{0}^{2}}\delta(t-t^{\prime})$, and $M$ is the
modal mass. Using Eqs.~(\ref{qdq}) and (\ref{u}) along with
Eq.~(\ref{Sq}), we obtain
\begin{equation}
\begin{aligned}
S_{q}(\omega)&=\frac{2}{\tau}\textrm{Re}\int_{0}^{\tau}dt\int_{0}^{t}dt_{1}\,u^{\ast}(t)u(t_{1})\exp\left[i(\omega-\omega_{0})(t-t_{1})\right]\\
&=\frac{2}{\tau}\textrm{Re}\int_{0}^{\tau}dt\int_{0}^{t}dt_{1}\int_{-\infty}^{t}dt^{\prime}\int_{-\infty}^{t_{1}}dt_{1}^{\prime}\,\Bigg\{
\exp\left[i(\omega-\omega_{0})(t-t_{1})-\Gamma(t-t^{\prime})-\Gamma(t_{1}-t_{1}^{\prime})\right]\\
&\shoveleft{\times\exp\left[-i\int_{t^{\prime}}^{t}\xi(t^{\prime\prime})dt^{\prime\prime}+i\int_{t_{1}^{\prime}}^{t_{1}}\xi(t_{1}^{\prime\prime})dt_{1}^{\prime\prime}\right]
\times\frac{\Gamma
k_{B}T}{M\omega_{0}^{2}}\delta(t^{\prime}-t_{1}^{\prime})\Bigg\}}\\
&=\frac{\Gamma
k_{B}T}{M\omega_{0}^{2}}\frac{2}{\tau}\textrm{Re}\int_{0}^{\tau}dt\int_{0}^{t}dt_{1}\int_{-\infty}^{t_{1}}dt^{\prime}\,\Bigg\{\exp\left[i(\omega-\omega_{0})(t-t_{1})-\Gamma(t+t_{1}-2t^{\prime})\right]\\
&\shoveleft{\times\exp\left[-i\int_{t_{1}}^{t}\xi(t^{\prime\prime})dt^{\prime\prime}\right]\Bigg\}}\\
&=\frac{k_{B}T}{M\omega_{0}^{2}}\frac{1}{\tau}\textrm{Re}\int_{0}^{\tau}dt\int_{0}^{t}dt_{1}\,\exp\left[i(\omega-\omega_{0})(t-t_{1})-\Gamma(t-t_{1})-i\int_{t_{1}}^{t}\xi(t^{\prime\prime})dt^{\prime\prime}\right]\,.
\end{aligned}
\label{Saveraged}
\end{equation}

In Eq.~(\ref{Saveraged}), we average over thermal noise since its
correlation time is short on the scale $\Gamma^{-1}\ll\tau$. We
can now separate $\xi(t)$ into two parts, one slow and the other
fast on the scale of the ring-down time $\Gamma^{-1}$:
$\xi(t)=\xi_{slow}(t)+\xi_{fast}(t)$. We assume that $\xi_{fast}$
is $\delta$-correlated, which allows to simplify the power
spectrum in Eq.~(\ref{Saveraged}) as
\begin{align}
S_{q}(\omega)&=\frac{k_{B}T}{M\omega_{0}^{2}}\frac{1}{\tau}\textrm{Re}\int_{0}^{\tau}dt\int_{0}^{t}dt_{1}\,\exp\left\{\left[i(\omega-\tilde{\omega}_{0})-\tilde{\Gamma}\right](t-t_{1})-i\xi_{slow}(t)(t-t_{1})\right\}\nonumber\\
&=\frac{k_{B}T}{M\omega_{0}^{2}}\frac{1}{\tau}\int_{0}^{\tau}dt\,\frac{\tilde{\Gamma}}{\tilde{\Gamma}^{2}+[\omega-\tilde{\omega}_{0}-\xi_{slow}(t)]^{2}}\,,\qquad
|\omega-\omega_{0}|\ll\omega_{0}\,,\label{Spaper}
\end{align}
where $\tilde{\Gamma}$ is the "instantaneous" half-width of the
spectrum and $\tilde{\omega}_{0}$ is the resonant angular
frequency, both renormalized by fast frequency noise. The relation
between $\Gamma$ and $\tilde{\Gamma}$ can be found in
Supplementary Ref.~\cite{Yaxing}. Equation~(\ref{Spaper}) is
Eq.~(2) in Methods. The resonant peak corresponding to
$S_{q}(\omega)$ only exists for
$\exp\left(-\tilde{\Gamma}\tau\right)\ll1$; the fact that
resonance lineshapes remain close to Lorentzian indicates that our
measurement times exceed $1/\tilde{\Gamma}$.

We emphasize that $\xi_{fast}(t)$ is $\delta$-correlated on the
"slow" time scale $\sim\Gamma^{-1}$, not on the fast scale $\sim
\omega_0^{-1}$. It is seen from Eq.~(\ref{Saveraged}) that the
components of $\xi(t)$ with frequencies much higher than $\Gamma$
are averaged out and therefore can be disregarded, as we indicated
earlier. The case where the frequency noise has significant
intensity near $2\omega_0$, so that it parametrically excites the
resonator, requires a separate analysis, but we have no
indications and no physical reasons to expect that strong
frequency noise with frequencies $\approx 2\omega_0$ is present in
our case.

The separation of the frequency noise into parts that are slow and
fast on the time scale $1/\Gamma$ leaves out a comparatively
narrow part of the noise spectrum. This part is averaged out when,
as in our case, the duration of a measurement $\tau\gg1/\Gamma$.
It is reasonable to expect that the contribution of this narrow
range of the frequency noise spectrum is small. Additional
information about the spectrum of the frequency noise can be
obtained by studying the power spectrum of the resonator in the
presence of periodic modulation \cite{Yaxing}.

The integral width of the spectrum can be obtained from
Eq.~(\ref{Spaper}). The integral width is expressed as the ratio
$\mathcal{I}=\mathcal{A}/S_{q}^{\textrm{max}}(\omega)$, where
$\mathcal{A}=\int_{0}^{\infty} S_{q}(\omega)d\omega$ is the
spectrum area and $S_{q}^{\textrm{max}}(\omega)$ is the resonance
height in the spectrum. The area $\mathcal{A}=\pi
k_{B}T/(M\omega_{0}^{2})$ is independent of frequency noise or
decay rate. For a Lorentzian spectrum ($\xi_{slow}=0$) we have
\[
\mathcal{I}^{-1}=\frac{1}{\pi\tilde{\Gamma}}=Q\frac{2}{\pi\tilde{\omega}_{0}}\,,
\]
where $Q$ is the quality factor.

Next, we calculate $\mathcal{I}^{-1}$, a good number to
characterize $Q$ for asymmetric and noisy resonances. For slow
frequency noise of weak intensity, the expression for the
reciprocal width takes on a "finite-time variance" form:
\begin{equation}
\mathcal{I}^{-1}\simeq\frac{1}{\pi\tilde{\Gamma}}\left(1-\frac{1}{\tilde{\Gamma}^{2}}\left[\frac{1}{\tau}\int_{0}^{\tau}dt\,\xi_{slow}^{2}(t)-\left(\frac{1}{\tau}\int_{0}^{\tau}dt\,\xi_{slow}(t)\right)^{2}\right]\right)\,,
\label{finite_time_variance}
\end{equation}
where $\frac{1}{\tau}\int_{0}^{\tau}dt\,\xi_{slow}(t)$ is the
measured shift of the resonant frequency resulting from slow
frequency noise. The term in square brackets in
Eq.~(\ref{finite_time_variance}) is always positive, resulting in
the two following effects:

\begin{enumerate}
\item{For a fixed measurement time $\tau$, multiple measurements of the integral width give an asymmetric distribution of $\mathcal{I}^{-1}$;}
\item{As $\tau$ increases, the peak value of the distribution of $\mathcal{I}^{-1}$ decreases, whereas the mean $\mathcal{I}$ increases. Once $\tau$ exceeds the noise correlation time,
$\mathcal{I}$ levels off to the value
$\pi\tilde{\Gamma}\left(1+\frac{1}{\tilde{\Gamma}^{2}}\langle\xi_{slow}^{2}(t)\rangle\right)$.}
\end{enumerate}

\begin{figure}[t]
\includegraphics{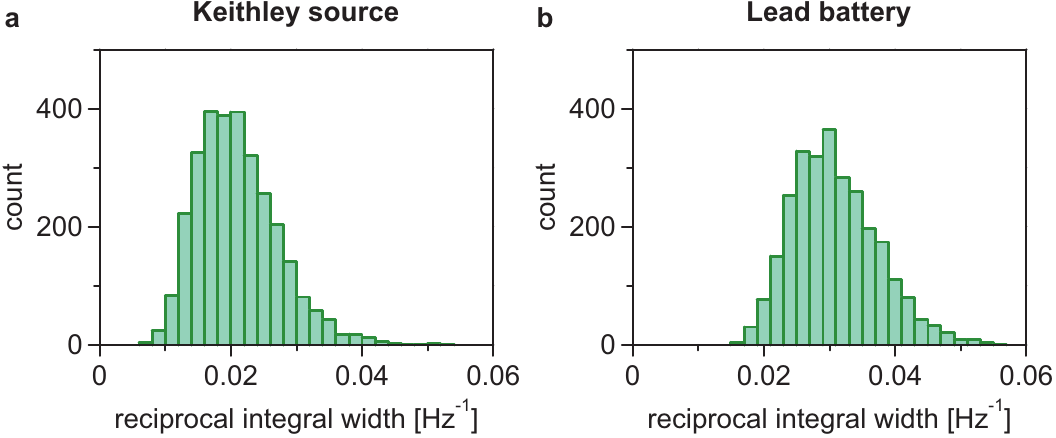}
\caption{Histograms of the reciprocal integral width
$\mathcal{I}^{-1}$, using the Keithley source (a) and the lead
battery (b) to bias the gate electrode.} \label{width}
\end{figure}
Supplementary Fig.~\ref{width} shows histograms of
$\mathcal{I}^{-1}$ built from the spectra used to construct the
histograms in Figs.~2c,d of the main text. If, as we assume, the
modulation of the conductance is linear in the resonator
displacement, the current-to-displacement scaling coefficient
drops out from $\mathcal{I}$. Therefore, the value of
$\mathcal{I}$ can be found directly from the data on the current
power spectra. Histograms of $\mathcal{I}^{-1}$ are asymmetric, in
agreement with effect~(1) predicted above. In addition, Fig.~3a of
the main text shows that the peak value of the linewidth
distribution tends to level off for long measurement times, in
agreement with effect~(2). (From that figure, we infer that the
spectrum of the frequency noise has a cutoff around 1/20~Hz.) Slow
fluctuations of the decay rate and/or the intensity of the thermal
noise would not lead to the observed increase of the the peak
value of the linewidth distribution with measurement time, but
rather would cause $\mathcal{I}$ to decrease.

\end{document}